 \documentclass[preprint2]{aastex}

\newcommand{\scs}{\scriptsize}
\newcommand{\til}{$\sim$}
\newcommand{\ergsqcmsec}{\thinspace\hbox{$\hbox{erg}\thinspace\hbox{cm}^{-2}
                \thinspace\hbox{s}^{-1}$}}
\newcommand{\ergsec}{\thinspace\hbox{$\hbox{erg}\thinspace\hbox{s}^{-1}$}}
\newcommand{\simlt}{\mbox{$^{<}\mskip-10.5mu_\sim$}}
\newcommand{\simgt}{\mbox{$^{>}\mskip-10.5mu_\sim$}}

\newcommand{\decsec}[2]{$#1\mbox{$''\mskip-7.6mu.\,$}#2$}

\newcommand{\decsectim}[2]{$#1\mbox{$^{\rm s}\mskip-6.3mu.\,$}#2$}

\newcommand{\asecbyasec}[2]{#1$''\times$#2$''$}
\newcommand{\aminbyamin}[2]{#1$'\times$#2$'$}

\def\today{\ifcase\month\or
January\or February\or March\or April\or May\or June\or
July\or August\or September\or October\or November\or December\fi
\space\number\day, \number\year}

\slugcomment{Accepted by the Astronomical Journal 2001 July 16}

\shorttitle{{\it Chandra} X-ray Position for the Rapid Burster in Liller 1}
\shortauthors{Homer et al.}

\begin{document}

\setcounter{footnote}{1}

\title{The Rapid Burster in Liller 1: the {\it Chandra} X-ray Position and a Search for an IR Counterpart$^{1,2}$\footnotetext{Based on observations with
the NASA/ESA
Hubble Space Telescope, obtained at the Space Telescope Science Institute,
which is operated by the Association of Universities for Research in
Astronomy, Inc., under NASA contract NAS5-26555.}\setcounter{footnote}{2}\footnotetext{Based on observations obtained with the Apache Point Observatory 3.5-meter telescope, which is owned and operated by the Astrophysical
Research Consortium, and on observations made with ESO Telescopes at the La Silla or Paranal Observatories under programme ID 51.5-0043.}
}

\author{L. Homer, Eric W. Deutsch\altaffilmark{3}, Scott F. Anderson}
\affil{Astronomy Department, Box 351580, University of Washington,
    Seattle, WA 98195-1580}
\email{homer@astro.washington.edu; deutsch@alumni.washington.edu;
       anderson@astro.washington.edu}
\and
\author{Bruce Margon}
\affil{Space Telescope Science Institute, 3700 San Martin Drive, Baltimore, MD 21218}
\email{margon@stsci.edu}


\received{2001 June 7}
\accepted{2001 July 16}


\altaffiltext{3}{Current address: Institute for Systems Biology, 4225 Roosevelt Way NE, Suite 200, Seattle, WA
98105-6099}

\begin{abstract}
Despite the unique X-ray behavior of the compact bursting X-ray source
MXB1730-335, the ``Rapid Burster" (RB) in the highly reddened globular cluster
Liller~1, to date there has been no known optical/IR counterpart for the object, no
precise astrometric solution that correlates X-ray, radio, and optical
positions and thus restricts the number of possible candidates, nor even
published IR images of the field. We solve a previous radio/X-ray positional discrepancy, presenting the results of precise {\it Chandra} X-ray
imaging, which definitively show that the radio source is positionally aligned with MXB1730-335.  At the same time, we have detected three
additional low luminosity ($L_X\sim10^{34}\ergsec$) X-ray sources within two core radii, which are possibly quiescent low-mass X-ray binaries.  We present both ground-based and {\it
Hubble Space Telescope} infrared imaging of the field (in quiescent and bursting X-ray states of the RB), together with the
necessary astrometric solution to overlay the radio/X-ray source positions. Even at {\it HST} resolution, the RB field is very
complex and there are multiple candidates. No object of unusual color, or
of substantial variability in quiescent versus active or burst versus
non-burst states, is identified.  Further, more sensitive {\it HST}/NICMOS and/or ground-based adaptive-optics observations are needed to confidently identify
the proper counterpart.  In the case of the RB, uncertain but plausible calculations on
the effects of the burst on the binary companion indicate that detection
of a variable candidate should be feasible.

\end{abstract}


\section{INTRODUCTION}

Unique in the Galaxy, the Rapid Burster (MXB 1730$-$335, hereafter RB) in the globular
cluster Liller 1 is the only low-mass X-ray binary (LMXB) star system
observed to emit both type~I and~II X-ray bursts (Hoffman et al. 1978).  Often the rapid
chains of bursts occur at intervals as short as 7~s, sometimes producing
thousands of bursts per day.  The burst flux is related to the time
between bursts in a fashion reminiscent of a relaxation oscillator,
strongly implying that the neutron star magnetic field acts as a reservoir
which fills and empties, causing each burst.  Thus with this object
we are directly observing the dynamic behavior of the intense magnetic
field of a neutron star on a timescale of seconds.  This activity only
occurs during the outburst phase, which occurs every 6-8 months and
lasts only a few weeks.  The {\it Rossi} X-ray Timing Explorer (RXTE)
All-Sky Monitor (ASM) has been continuously monitoring the X-ray sky for years, and has to date observed many outburst
episodes from the RB.  This outburst behavior is described
in detail by Guerriero et al.~(1999), and a more general summary of the
properties of the RB is given by Lewin et al.~(1995).

Despite considerable study of this object in X-rays, there remains no
optical counterpart, which is almost surely due to the extreme crowding
from cluster stars as well as high interstellar extinction.   There have
been past searches for IR bursts, analogous to the extraordinary X-ray
behavior, using large ($\sim20''$) apertures centered on the cluster.
The results, summarized by Lewin et al.~(1995), are tantalizing but
ambiguous.  Direct identification of a stellar IR counterpart would
thus be extremely important in planning of follow-on studies of this
fascinating object.  Optical work will be extremely difficult for
this system, as the interstellar reddening is E$(B-V)=3.0$ for Liller 1
(Frogel et al.~1995; Ortolani et al.~1996).  In passing we note that the
cluster itself is also astrophysically important: Frogel et al.~(1995)
conclude it is the most metal rich cluster known.

The most accurate X-ray position to date for the RB was derived from {\it
Einstein} observations by Grindlay et al.~(1984), but the uncertainty
there is still several arcseconds.  Moreover, there is a considerable
($\sim5\sigma$) discrepancy between that X-ray position and recently determined positions of a probable radio counterpart.  Moore et al.~(2000) and  Fruchter \& Goss (2000) both report on a variable radio source near the center of the
cluster. Specifically, Moore et al.~(2000) found an apparent correlation between the changes in radio and X-ray fluxes, which strongly implies that the radio counterpart to the Rapid
Buster has indeed been isolated.  These authors also discuss in detail a likely explanation for the X-ray/radio positional discrepancy,
wherein the uncertainty in the X-ray position is underestimated and the
uncertainty distribution strongly non-Gaussian.

In order to both resolve the positional discrepancy and determine a far more accurate X-ray position, we have obtained high angular resolution
X-ray imaging of the Liller 1 field with {\it Chandra X-ray Observatory} and HRC-I. Working within a much smaller error circle for the X-ray/radio source, we have
made the first IR/optical imaging search for the counterpart in these bands.  We make use of both ground-based observations
obtained during an outburst episode and {\it Hubble Space Telescope}
observations obtained during a quiescent phase.  Furthermore, we have serendipitously detected three low luminosity X-ray sources in the core, for which
we also analyze the {\it HST} IR photometry.

\section{X-ray data analysis and results}

\subsection{{\it Chandra} HRC-I observations}
We observed Liller 1 with {\it Chandra} for 15.4 ks on 2000 August 2.  The \til\aminbyamin{30}{30} field-of-view high resolution
camera imager (HRC-I; Murray et al. 1997) was approximately centered on the cluster center.  This mode provides the highest possible spatial resolution
available from {\it Chandra}, to achieve our primary science goal, and its large FOV raises the probability of detecting additional X-ray
sources, which can be matched to cataloged optical star positions, thus improving the astrometry even farther.  However, unlike the X-ray CCD detectors,
the HRC-I detector has very poor intrinsic energy resolution, so we have no spectral information.  On the other hand, it does provide very
good  time-resolution, down to 16$\mu s$.

\subsection{X-ray position of the Rapid Burster}
Data reduction was undertaken with routines in {\tt CIAO v2.1}.  In addition to the very bright RB, a Mexican-Hat wavelet source detection routine ({\tt wavdetect}) found 9
sources at greater than 3$\sigma$ above background in the entire field, beyond  the RB's contaminating PSF wings (a \til10\arcsec radius region).  The routine also
gave the source positions, according to the well-determined satellite
aspect for this observation.  We then compared all
these X-ray source positions with both the USNO A-2 and Tycho-2  catalogs  (Monet et al. 1998; H{\o}g et al 2000) which contain very accurate (within \decsec{0}{3} and \decsec{0}{06}
respectively) star positions, to identify any matches.  Unfortunately, only one match was made to a 10th magnitude star, TYC 7380-976-1, in the Tycho-2 catalog.  An
offset of  \decsec{0}{56} and \decsec{0}{08} in R.A. and dec. was found between its nominal {\it Chandra} and cataloged positions.  This is
consistent at $\simlt2\sigma$ level with the expected uncertainty of \til \decsec{0}{4} of the {\it Chandra} aspect solution for the HRC-I, as determined for many such X-ray-optical pairs in all
observations to date.  We therefore apply this correction to the nominal {\it Chandra} position for the RB, and derive a new X-ray position of
$\alpha=17^h33^m$\decsectim{24}{61}$\pm$\decsectim{0}{03}, $\delta=-33^{\circ}23'$\decsec{19}{90}$\pm$\decsec{0}{4}, where we conservatively
quote our uncertainties as those of the {\it Chandra} aspect.

\begin{figure}[!htb]
\plotone{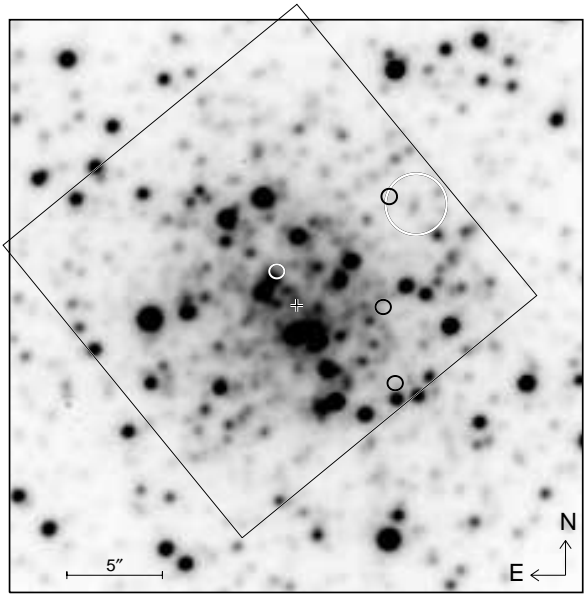}
\caption{\asecbyasec{30}{30} archival NTT $z$-band image (Ortolani,
Bica, \& Barbuy 1996) of Liller 1, obtained with \decsec{0}{5} seeing.
The tilted square indicates the full \asecbyasec{30}{30} {\it HST} NICMOS field of
view. A cross is drawn at the cluster center.  The small circles ($1\sigma$) denote the positions of the four {\it Chandra} X-ray sources; that of the Rapid
Burster in white.  This is clearly inconsistent with the
location of the much larger {\it Einstein} X-ray error circle, also shown in white. (It is curious to note that one of our low luminosity {\it
Chandra} sources is aligned with the old {\it Einstein} position). \label{fig:posns}}
\end{figure}

\begin{figure*}[!htb]
\resizebox*{!}{0.95\textwidth}{\rotatebox{0}{\plotone{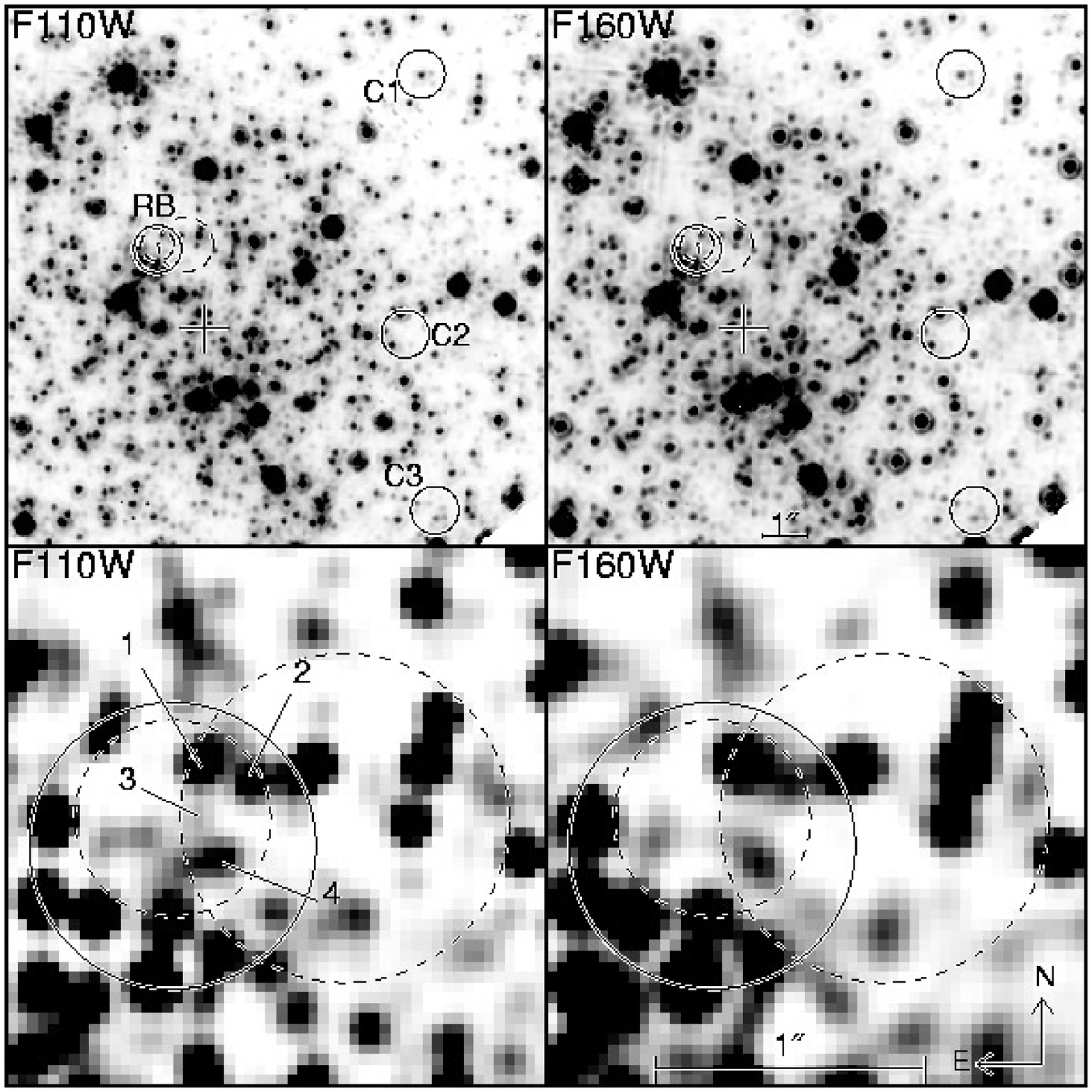}}}
\caption{{\it HST} NICMOS images (FWHM
$\sim$\decsec{0}{1}) of the center of the globular cluster Liller 1, through the F110W (left) and
F160W (right) filters.  Upper panels: \asecbyasec{14}{14} field of view is shown encompassing the positions of the Rapid
Burster (RB) and the three new low luminosity sources detected by {\it Chandra} (C1-3). The $1\sigma$ error regions for the positions are overlaid: the solid circles are the {\it Chandra} X-ray results, and the 
dashed circles are the two independent radio determinations for the Rapid
Burster. Lower panels:  a zoom-in (\asecbyasec{2}{2} field) on the Rapid
Burster position.  Note how complex the field is, even within the most probable overlap region four stars are visible; these are labelled for cross-reference
to the color-magnitude diagram (fig.~\ref{fig:CMD}\label{fig:posns2}).}
\end{figure*}

We are then immediately able to compare the new {\it Chandra} X-ray position to the various published positions, as tabulated in
table~\ref{tab:posns} and illustrated in figures~\ref{fig:posns} and \ref{fig:posns2}. There can be no doubt about the consistency of the new X-ray position with the
two radio results, providing compelling evidence of their association, and thus confirming the inaccuracy of the old {\it Einstein} result.

\begin{table*}[!htb]
\caption{X-ray and radio positions for the Rapid Burster (in the ICRS).\label{tab:posns}}
\begin{center}
\begin{tabular}{l l l l l l l} 
\tableline\tableline
Band & Telescope &$\alpha$ (J2000) & Error & $\delta$ (J2000) & Error & Reference\\
&&(h, m, s)& ($1\sigma$)&($^\circ$, \arcmin, \arcsec)&($1\sigma$)&\\
\tableline
X-ray & {\it Chandra}/HRC-I & 17 33 24.61 & \decsectim{0}{03} & -33 23 19.9 & \decsec{0}{4}& current work\\
X-ray &{\it Einstein}/HRI& 17 33 24.09 & \decsectim{0}{13} & -33 23 16.4 & \decsec{1}{6}& Grindlay et al. 1984\\
Radio & VLA (D$\rightarrow$A) & 17 33 24.61 &\decsectim{0}{01}& -33 23 19.8 & \decsec{0}{1}&Moore et al. (2000)\\
Radio & VLA (CnB)  & 17 33 24.56  &\decsectim{0}{03} & -33 23 19.8 & \decsec{0}{5}&Fruchter \& Goss (2000)\\
\tableline
\end{tabular}
\end{center}
\end{table*}

\subsection{X-ray state of the Rapid Burster}
\begin{figure*}[!htb]
\resizebox*{.95\textwidth}{0.45\textheight}{\rotatebox{-90}{\plotone{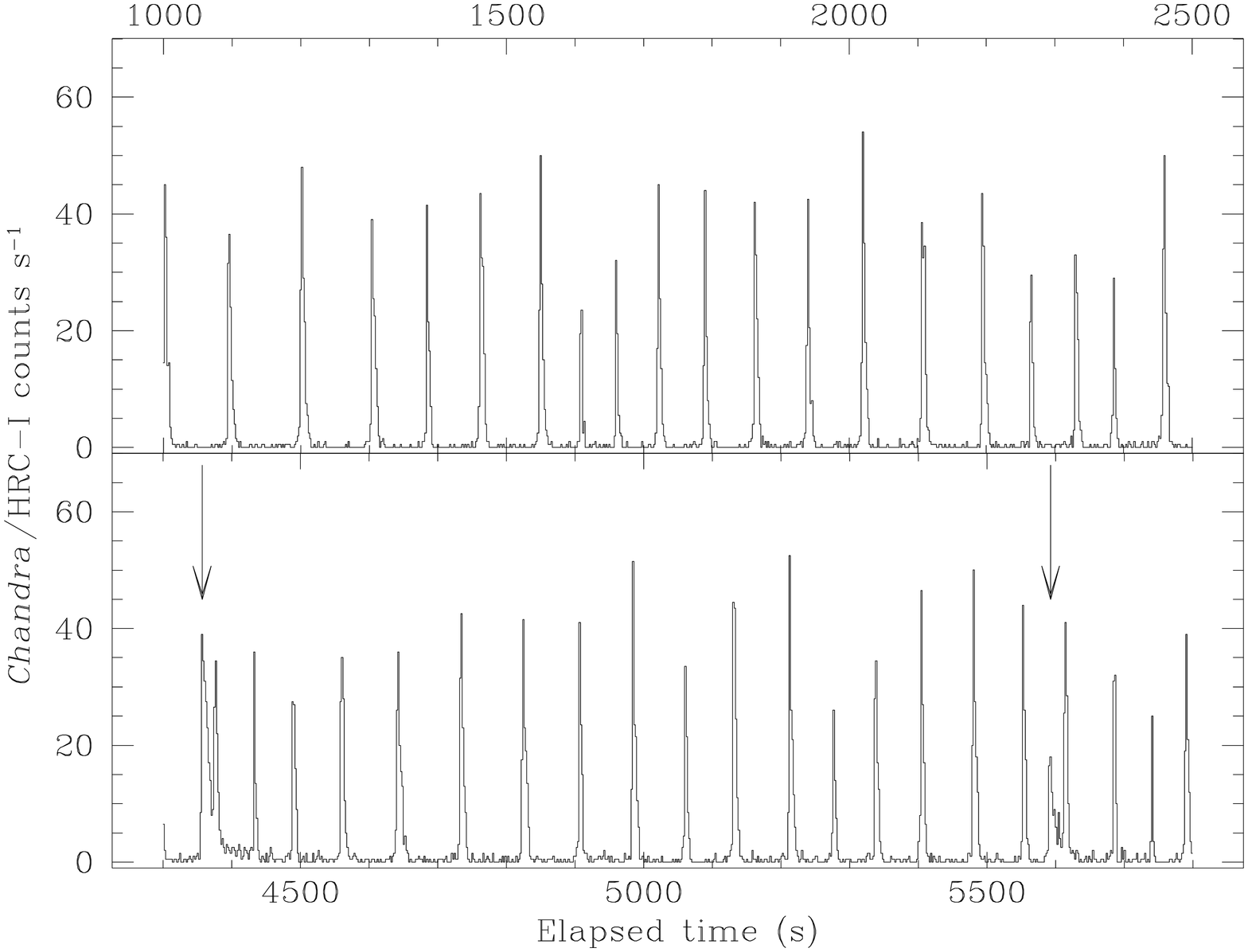}}}
\caption{Illustrative 1500s-long sections of the {\it Chandra}/HRC-I lightcurve of the Rapid Burster, with 1s binning applied.  Throughout the
\til13 ks observation the source shows regular type-II bursts (top panel), but in addition we detect two type-I burst event (bottom panel,
indicated by arrows).\label{fig:CXOlc}}
\end{figure*}

\begin{figure*}[!htb]
\resizebox*{.95\textwidth}{!}{\plotone{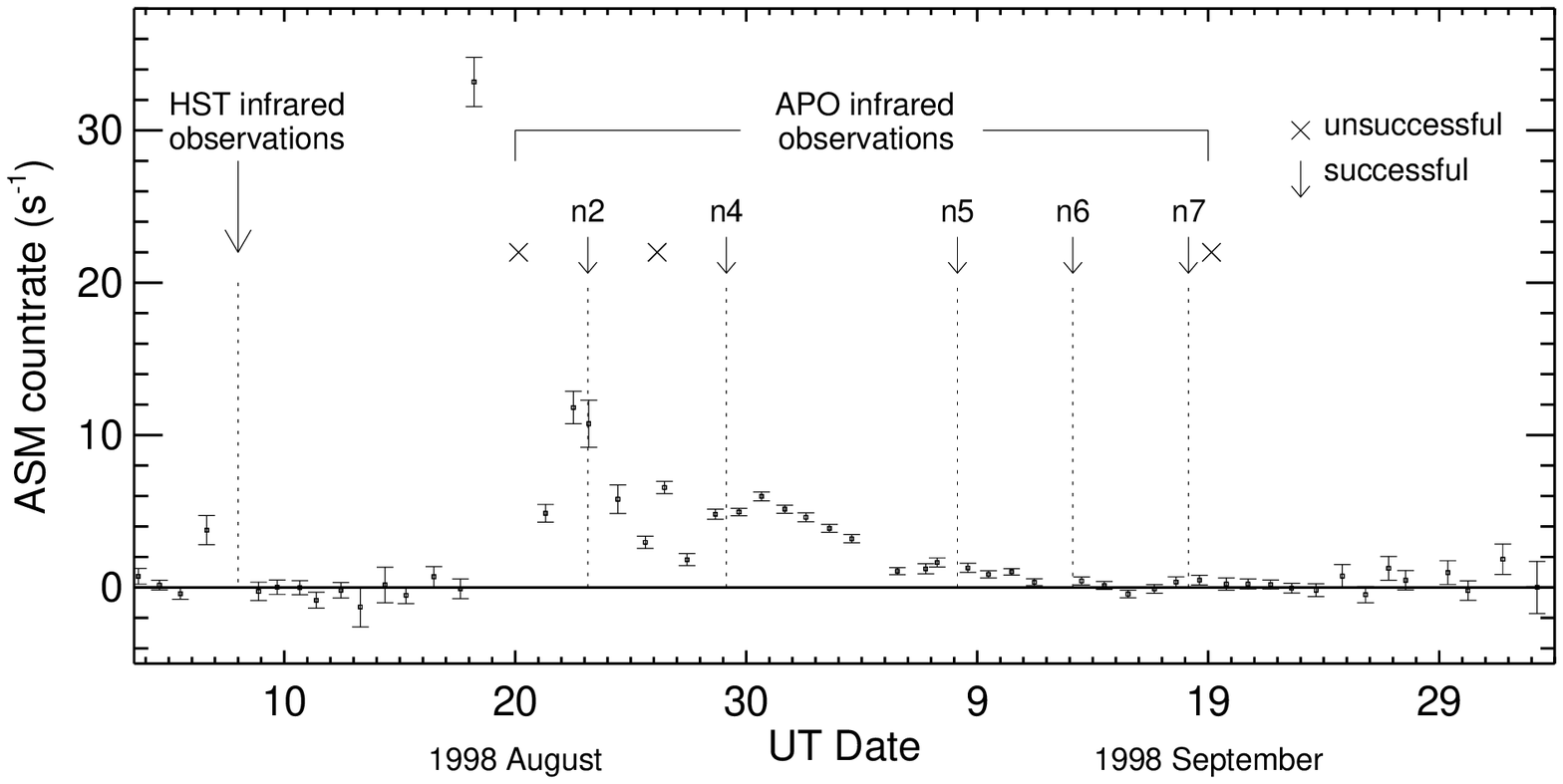}}
\resizebox*{.95\textwidth}{!}{\plotone{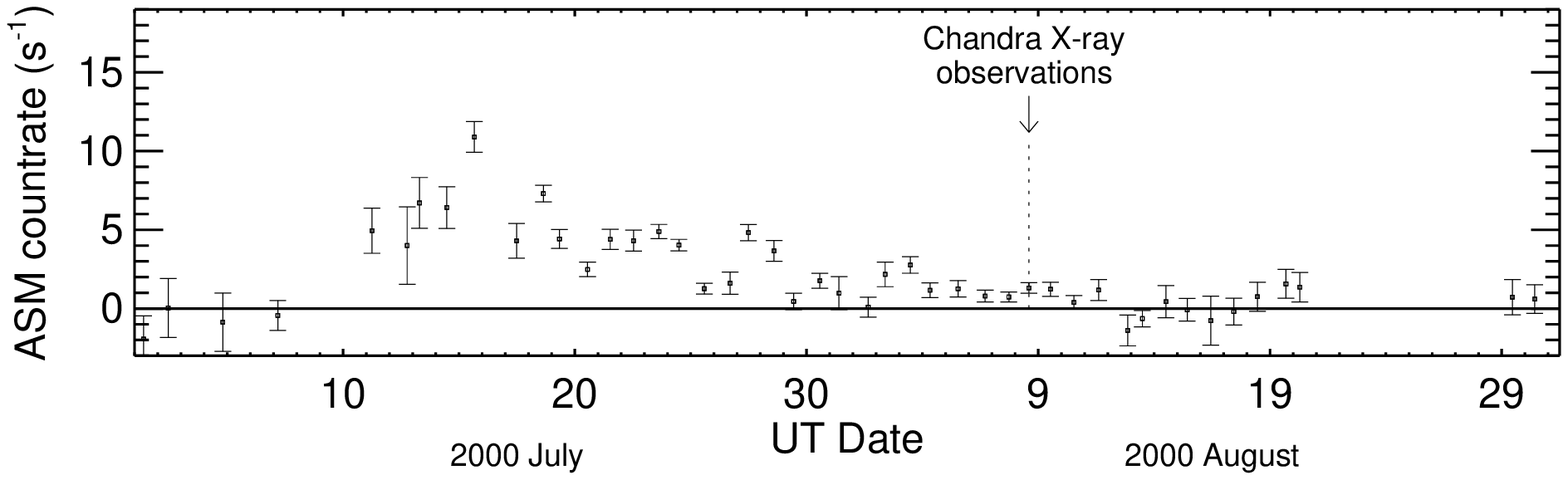}}
\caption{Relevant sections of the X-ray light curve of the Rapid Burster from the {\it
Rossi} X-ray Timing Explorer ({\it RXTE}) All-Sky Monitor (ASM). One day average count rates have been shown for clarity. Upper panel: the epochs
of our infrared observations have been overlaid.  The bright X-ray phase of the
source, which accompanies the rapid bursts clearly begins around 1998
August 17.  The {\it HST} IR observations were made in the quiescent period ahead of the outburst, in contrast to the ground-based IR observations
from APO, which were made 
during the outburst.  Lower panel: here we have marked the epoch of our {\it Chandra} HRC-I X-ray observation,
which also caught the tail end of an X-ray bright phase starting in early 2000 July, consistent with the active bursting behaviour shown in
figure~\ref{fig:CXOlc}.\label{fig:ASMlc}}
\end{figure*}

An X-ray lightcurve was also extracted using all available data.  A circular region of 2\arcsec\ radius  (encircling 97\% of the energy) and an annulus
(80\arcsec--120\arcsec) were defined centered on
the RB to measure source flux and background respectively.  The events were then summed into 1s bins using the {\tt CIAO} routine {\tt
lightcurve}, for each region.   Given the brightness of the source flux, and the steady very low level of background
flux, we decided that background subtraction was unnecessary.  Two illustrative sections of the lightcurve are shown in figure~\ref{fig:CXOlc}; showing
type-II bursts every \til100s, and two additional type-I burst events (judging solely on
morphology) in the lower panel. In terms of the evolution of RB outbursts (Marshall et al 1979; Guerriero et al. 1999), the source was clearly in an active state,
exhibiting the rapid, quasi-regular type-II bursting characteristic of the final stages (mode-II) of the phase II of an outburst.  We also note
that the level of the persistent (i.e. between all bursts) flux did decrease by \til30\% over the course of the observation. Consideration of the ASM
lightcurve (figure~\ref{fig:ASMlc}, lower panel) confirms that our {\it Chandra} observation was indeed made in the tail of an outburst, which began
about July 10th, 29 days previous.   In general, this July 2000 outburst appears to have been weaker than either that of August 1998 or
others reported by Guerriero et al. (1999).  We have used the range of best fit
two-component blackbody spectra as measured by these authors, to convert HRC-I counts to flux, and indeed we measure the persistent unabsorbed
flux (0.2--10 keV) to be $3-6\times10^{-11}$ \ergsqcmsec, more than a 1000 times fainter than quoted by Guerriero et al. at a similar stage
during other outbursts.  This is yet another example of the complex outbursting behaviour
of the RB.  We do note, however, that low as this flux is, it is still \til10 times brighter than the quiescent
detection made by {\it ASCA} (Asai et al. 1996).

\subsection{Serendipitous low luminosity X-ray sources in the core}
The exquisite PSF of {\it Chandra}/HRC-I opens up for the first time the possibility of detecting a population of fainter X-ray sources
co-inhabiting the center of a cluster with a bright LMXB.  We have undertaken such a search within the half-mass radius (27\arcsec, Harris 1996) of Liller 1.    This region is
still largely dominated by the wings of the RB PSF.  However, using {\tt makegti} and {\tt dmfilter} routines we have created a summed image,
using only time-intervals when the RB is between bursts, and many times fainter.   We used a sliding cell technique ({\tt celldetect}) to identify sources, which were then centroided to
give the nominal {\it Chandra} positions.  Four sources were detected; the RB of course, plus two additional ones at $\simgt$5 $\sigma$
confidence and a third at \til4 $\sigma$.  The corrected positions, count rates, assumed spectral models and
corresponding luminosities (quoted in the 0.5--2.5 keV band for ease of comparison to {\it ROSAT} results) are detailed in table~\ref{tab:CXO_dim}.  All three low luminosity sources lie within approximately two core radii, taken as
\decsec{3}{6} (Harris 1996), a region in which we would only expect $\sim1\times10^{-3}$ extragalactic background sources at these flux
levels or greater (Giacconi et al. 2001), so we can be reasonably confident of their association with
the cluster.  

Indeed, the relatively high $L_X$ values of our new sources are immediately suggestive as to their nature.  The sensitive {\it
Chandra}/ACIS spectro-imaging surveys of the globular clusters 47 Tuc (Grindlay et al. 2001) and $\omega$ Cen (Rutledge et al. 2001) have
revealed that the brightest sources are
most likely quiescent LMXBs (qLMXBs) on X-ray spectral grounds, with the next brightest class, probably magnetic CVs, 10-100 times fainter.  These
new Liller 1 sources have $L_X$ values consistent with these qLMXBs.

\begin{table*}[!htb]
\begin{minipage}{12cm}
\caption{Positions (in ICRS) and X-ray properties for all sources detected by  {\it Chandra} in the core of Liller 1.\label{tab:CXO_dim}}
\begin{tabular}{l l l l l l} 
\tableline\tableline
{\scs Object }&{\scs Position }&{\scs Error }&{\scs HRC-I count rate }&{\scs Spectral
model }&{\scs $L_X$ (.5-2.5 keV)}\\
& {\scs $\alpha$, $\delta$ (J2000)}&{\scs ($1\sigma$)}&{\scs (counts s$^{-1}$)}&{\scs (blackbody or }& {\scs (\ergsec)}\tablenotemark{a}\tablenotetext{a}{Luminosity for
a distance of 10.5 kpc, and corrected for substantial interstellar absorption by
$N_H=1.7\times10^{22}$cm$^{-2}$ (Marshall et al. 2001)}\\
&{\scs(h, m, s), ($^\circ$, \arcmin, \arcsec)}&&&{\scs thermal bremsstrahlung)}\tablenotemark{b}\tablenotetext{b}{The range of $kT_{BB}$ for the RB is that found by Guerriero et al. (2000) for the dominant cooler
blackbody component in their fit to its persistent emission.  For the other sources representative temperatures are shown for blackbody fits to
qLMXBs, and thermal bremsstrahlung fits to CV spectra.}&\\
\tableline
{\scs Rapid Burster}&{\scs  17 33 24.61}\tablenotemark{c}\tablenotetext{c}{Position as derived from all data, but that for the
persistent emission alone is fully consistent.}&{\scs \decsectim{0}{03} }&{\scs $0.345\pm0.005$}\tablenotemark{d}\tablenotetext{d}{persistent emission}&{\scs $kT_{BB}=0.25$keV}&{\scs $6.4\times10^{34}$}\\
&{\scs -33 23 19.9}&{\scs \decsec{0}{4}} &&{\scs $kT_{BB}=0.4$keV}&{\scs $3.4\times10^{35}$}\\

{\scs CXOU J173324.1-332316.0  }&{\scs 17 33 24.14 }&{\scs\decsectim{0}{03}}&{\scs $1.12\pm0.45\times10^{-3}$ }&{\scs $kT_{BB}$=0.3 keV }&{\scs $2\times10^{33}$}\\ 
{\scs (C1)}&{\scs -33 23 16.0 }&{\scs \decsec{0}{4}}&&{\scs $kT_{BB}$=0.1 keV }&{\scs
$4\times10^{34}$}\\
&&&&{\scs $kT_{Brems}$=5 keV }&{\scs $9\times10^{32}$}\\ 

{\scs CXOU J173324.1-332321.7  }&{\scs 17 33 24.17 }&{\scs\decsectim{0}{03}}&{\scs $1.27\pm0.47\times10^{-3}$ }&{\scs $kT_{BB}$=0.3 keV }&{\scs $2\times10^{34}$}\\ 
{\scs (C2)}&{\scs -33 23 21.8 }&{\scs \decsec{0}{4}}&&{\scs $kT_{BB}$=0.1 keV }&{\scs
$5\times10^{34}$}\\
&&&&{\scs $kT_{Brems}$=5 keV }&{\scs $1\times10^{33}$}\\ 

{\scs CXOU J173324.1-332325.7  }&{\scs 17 33 24.12 }&{\scs\decsectim{0}{03}}&{\scs $0.92\pm0.41\times10^{-3}$ }&{\scs $kT_{BB}$=0.3 keV }&{\scs $1\times10^{33}$}\\ 
{\scs (C3)}&{\scs -33 23 25.7 }&{\scs \decsec{0}{4}}&&{\scs $kT_{BB}$=0.1 keV }&{\scs
$3\times10^{34}$}\\
&&&&{\scs $kT_{Brems}$=5 keV }&{\scs $7\times10^{32}$}\\ 

\tableline
\end{tabular}
\end{minipage}
\end{table*} 

\section{Search for IR counterparts}

\subsection{Optical/IR Astrometry}

In general, the very accurate USNO-A2.0 star catalog (Monet et al. 1998) makes it
possible to tie an arbitrary field rather easily to the International
Celestial Reference System (ICRS) with sub-arcsecond precision.  We use
this catalog to tie the field of Liller 1 to the ICRS, and use the
empirical uncertainty estimators of Deutsch (1999) to determine the
size of the error circles we examine.  We adopt
an optical $1\sigma$ radial uncertainty of \decsec{0}{35} from the last
row of Table 1 in that work.

We begin with a $z$-band image of Liller~1 obtained from ESO NTT
science archive; this \asecbyasec{130}{130} image was first discussed
in a study of Liller~1 by Ortolani et al. (1996).  The stellar profiles
have FWHM \decsec{0}{50}, and the image is sampled at \decsec{0}{1288}
pixel$^{-1}$.  We select 33 bright and isolated stars in common between
the USNO-A2.0 catalog (epoch 1975.4 in this field) and the $z$-band image
(epoch 1994.4) and fit an astrometric solution to the image using IDL
procedures written by E.W.D. and from the {\it Astronomy User's Library}
(Landsman 1993).  The residuals of the fit ($\sigma$=\decsec{0}{17})
imply an approximate uncertainty ($\sigma/\sqrt{n-3}$) in the alignment
to the USNO-A2.0 frame of \decsec{0}{03} before considering proper
motion effects.

In Fig.~1 we show a \asecbyasec{30}{30} region of the $z$-band image
centered on the cluster.  We have estimated the approximate center of the cluster from this image as marked with a $+$ symbol.  The 1 $\sigma$ error
regions for X-ray sources are also overplotted; the respective positions for the RB in white, from {\it Einstein} (large) and {\it Chandra}
(small), whilst the three low luminosity sources detected by {\it Chandra} are in black.  The tilted square indicates the
field-of-view of the {\it HST} IR image, the central part of which is shown in more detail in Fig.~\ref{fig:posns2}.

\subsection{{\it HST} NICMOS Imagery}

There are many {\it HST} NICMOS images of Liller 1 in the {\it Hubble
Data Archive}, but only one set of observations are positionally suitable for a search
for the RB in the core of the cluster.  The two observations
are made with the NIC2 camera (\asecbyasec{19}{19} field of view sampled
at \decsec{0}{075} pixel$^{-1}$) and the F110W and F160W filters, which
are similar to the $J$ and $H$ bands, respectively.  We reprocess the
raw n4lc01010 and n4lc01040 datasets, observed on 1998 August 8, with
the STSDAS {\it calnica} task and the latest calibration reference files.

Each observation includes four individual exposures at different offsets.
Since the field is so crowded, even by {\it HST} standards, we employ
the STSDAS {\bf dither} package (Fruchter et al.~1998) to combine the
four frames on a finer pixel grid.  The frames are drizzled into a
final image with \decsec{0}{0378} pixels.  This process significantly
mitigates the loss of resolution due to the undersampling of the PSF. In Fig.~\ref{fig:posns2} we display the resulting combined images.  The images
were obtained with the F110W (left) and F160W (right) filters, and the central peaks of
stellar objects have FWHM \decsec{0}{11} and \decsec{0}{13}, respectively.  The upper panels show a \asecbyasec{14}{14} field including the positions of all 4
sources detected by {\it Chandra}.  The lower panels show restored
images, zoomed-in on the RB position alone; we applied a Lucy-Richardson restoration technique, in
order to remove the pronounced Airy ring structure in the wings of the PSF.

On each frame we draw three sets of error regions.  The small solid circles are those of our {\it Chandra} X-ray source
positions, whilst the smaller
dashed circle denotes
the RB radio position of Moore et al.~(2000) and larger dashed circle denotes the RB radio position of Fruchter \& Goss (2000).  In each case the
size of the
axes represents the combination of the optical astrometric uncertainties (\decsec{0}{35}), with the $1\sigma$ X-ray/radio uncertainties (see
tables.~\ref{tab:posns} and ~\ref{tab:CXO_dim}). The center of the cluster is again marked.

\begin{figure}[!htb]
\resizebox{.495\textwidth}{!}{\rotatebox{0}{\plotone{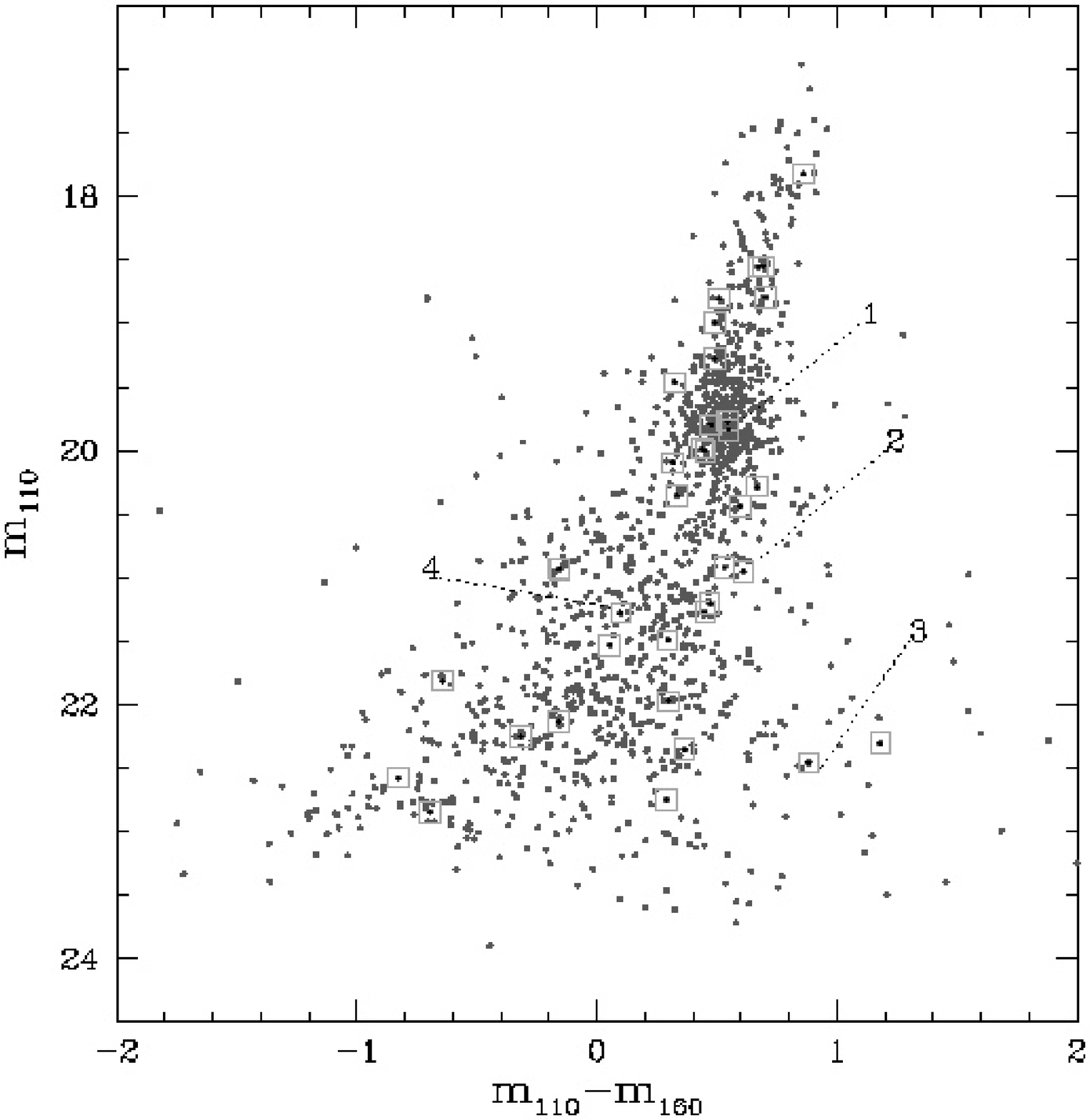}}}
\caption{IR color-magnitude diagram for the central region of Liller 1, encompassing the fields of the detected X-ray (and radio) sources.  The boxes indicate the stars which lie within each error
region.  The numbered stars are those within the overlap of the X-ray/radio error regions for the Rapid Burster (see fig.~\ref{fig:posns2}).  Within the uncertainties of
the photometry, none of the stars of positional interest can be said to show anomalous colors. \label{fig:CMD}}
\end{figure}

We have performed aperture photometry on all stars within 2\arcsec\ of each X-ray/radio source position, and constructed an IR  color-magnitude
diagram (see fig.~\ref{fig:CMD}).  The stars within each 1 $\sigma$ error region are identified by the boxes.  In the
case of the RB, its counterpart is most likely to lie within the overlap region of its three positional determinations, hence we have also numbered four stars therein.  Unfortunately, it is
immediately obvious than none of these stars of positional interest have markedly unusual IR colors at the epoch of the {\it HST} observations.  As the RB was in quiescence during this NICMOS observation (see fig.~\ref{fig:ASMlc}), all four
sources were X-ray faint at
the time.  The lack of unusual IR colors implies that there was little contribution from any accretion disc and that we are mostly
observing the secondary star in each case; hence their IR colors are those of a typical cluster star.  In terms of variability, a qLMXB typically shows
ellipsoidal variations, due to the changing aspect of the distorted (Roche lobe-filling) donor star.  However, these observations were
relatively short, and can place no useful constraints on such variability.

\subsection{Apache Point K-band Observations}

In an attempt to detect the RB counterpart by observing its IR variability, we have also obtained a set of time-resolved, $\sim1''$ resolution
imaging observations of Liller 1 in the K~band, taken with the Apache
Point (APO) 3.5-m telescope during a recent X-ray outburst episode which began
around 1998 August 17 (Fox \& Lewin 1998).  The observations were acquired
on multiple nights in 1998 August-September, approximately one hour per night,
including times of simultaneous pointed X-ray observation by RXTE during
which X-ray bursts were recorded.  We search both the X-ray and radio
error circles for a variable object, using an image-subtraction technique.
The infrared APO observations were done with the GRIM II infrared imager, which yields \decsec{0}{237}
pixel$^{-1}$ over the $256\times256$ array in its F/10 mode.  We
attempted observations on eight nights, of which five yielded usable
data.  Figure~\ref{fig:ASMlc} shows the ASM X-ray light curve with the epochs of our IR
APO observations overlaid.  Due to the hour angle and southern
declination, only $1-2$ hr of observation was possible on each night.
The best exposures were stacked to create a deep image for each night.
The resolution of the stacked images ranged from \decsec{0}{8} to
\decsec{1}{6}.  GRIM often produces better image quality than this,
but not at this large airmass $(z>2.5$).

The IR images are stacked to improve signal to noise and remove detector
imperfections.  First we look for possible variability between nights
by stacking the best exposures from each night and subtracting the stacks
from one another.  The stacked images are rotated and shifted to align
optimally, and the image with the better PSF is convolved with a kernel
which causes both images to have matching PSFs.  The kernel is
determined with the {\it psfmatch} task in the IMMATCH package within
IRAF.  Finally, the two images are scaled to the same relative flux and
subtracted.

No obvious significant flux difference is evident within the X-ray/radio error circles
in the subtracted images.  To quantify the flux difference that should
have been detectable, we add artificial stars in the original stacked
images.  We extract a bright star PSF, scale it appropriately, and add it
to the images in regions near the error circles with similar flux levels.
We find that we could have easily detected a flux
difference corresponding to a $K=14$ object fading by 1 mag.

On 1998 August 28 one orbit of {\it RXTE} PCA X-ray observation was
simultaneous with our IR observation.  Type I and II X-ray bursts were
detected during this X-ray observation.  The X-ray observation window
and the observed burst times were kindly provided to us by D. Fox
(private communication).

We stack all the IR exposures which occurred during an X-ray burst and
subtracted a stack of exposures which were obtained between bursts.
Again, no obvious excess flux is detected.  Using artificial star
tests, we determine that we could have easily detected a flux
difference corresponding  to a
$K=14$ object fading by 1 mag inside the X-ray/radio error circles.

\section{DISCUSSION}
The accurate X-ray position and corresponding astrometric solution imply that there is unlikely to be a confident identification of an IR
counterpart to the RB on positional grounds alone: the field is simply too complex, and we are largely ignorant of the expected IR properties of this
exotic object.  Thus detection of variability might be a key to identification of the counterpart.

Previous IR
observations of this object made use of large aperture ($\sim20''$)
photometers.  Several IR bursts were claimed in some of these previous
studies (e.g. Kulkarni et al. 1979) although additional extensive IR
non-detections are documented by Lawrence et al. (1983a), who cast
doubt upon previous claims.  Indeed, the bursts detected by Kulkarni et
al. (1979) would be fabulously bright if real: the peak intensities
appear to be comparable to the total flux of the center $20''$ of the
entire cluster.  Such intense bursts would have nearly
saturated our detector.

In order to make a separate estimate of the IR flux we might expect, we
use the example of an optical burst observed from MXB\,1636--53 by
Lawrence et al. (1983b).  Using the observed persistent and brightest UBV
magnitudes for that object, we estimate a persistent $K=16.3$ and
brightest $K=15.7$ based on the 25,000 and 50,000 K blackbody
temperatures, respectively, derived by Lawrence et al. (1983b).
Assuming a similar distance and correcting for the reddening to
Liller~1, E$(B-V)=3.0$, we crudely estimate a persistent $K=17.2$ and
brightest $K=16.6$.  This argument was originally put forward by Fox et al. (1998) who reached a similar conclusion.
It is unclear how accurate these estimates may be, as
the absolute optical/IR luminosity depends heavily upon the size of the
reprocessing region; moreover as
not even an orbital period is known for the RB, we have no idea whether its geometry is comparable to MXB\,1636--53.  Our observed upper limits
from the ground-based IR data 
fall short of testing these model estimates, principally due to seeing
effects.  However, such tests should be feasible with future {\it HST} or ground-based adaptive-optics IR observations.

In the case of the three probable qLMXBs we have detected with {\it Chandra}, the identification of counterparts may prove even more
difficult. Detection of ellipsoidal variability or another observation catching a source in outburst might work.  However, we have little
knowledge as to their outburst recurrence timescales, as X-ray source confusion with the RB has always been a limitation before. Even the
long-term coverage of the Liller~1 field by {\it RXTE}/ASM shows no evidence for more than one source going into outburst, and pointed
observations during these outbursts have always confirmed the RB as the source.  Interestingly, one of low luminosity sources, CXOU J173324.1-332316.0, is aligned  with the old {\it Einstein} position
for the RB.  One might speculate that if this source were in outburst at the time of that determination it might have contributed at least to
an underestimation of the true positional uncertainties for the RB. In any case our new precise {\it Chandra} position, consistent with two
independent radio positions, confirms without doubt the true location of the RB within Liller 1.

\acknowledgments

We thank Derek Fox and his collaborators for sharing {\it RXTE} PCA burst
timing information. {\it RXTE} ASM data
products were provided by the ASM/RXTE teams at MIT and at the {\it RXTE}
SOF and GOF at NASA's GSFC.  We are grateful to the anonymous referee for his helpful comments. Support for this work was provided by NASA
through grants NAG5-7330 and NAG5-7932, as well as grant AR-07990.01
from the ST\,ScI, which is operated by AURA, Inc., and SAO grant GO 0-011x.


\vspace{5mm}
{\em Note added in proof} -- A. Tennant (2001, private communication) has derived an independent position for the RB from {\it Chandra} timing
data, which while not as accurate as that discussed here, does tend to confirm the positional association of the X-ray and radio sources.

\end{document}